\documentclass[12pt]{article}
\usepackage{array,verbatim,amssymb,amsbsy,fancyhdr,mathrsfs}
\usepackage{citesort}
\usepackage{mathrsfs}
\usepackage{stmaryrd}
\usepackage{textcomp}
\usepackage[dutch]{babel}
\usepackage{makeidx}
\usepackage{latexsym}
\usepackage{epsfig}
\usepackage{amsmath}

\renewcommand{\thefootnote}{\fnsymbol{footnote}}

\begin{document}
\sloppy
\vspace*{2cm}
\noindent
{\bf Translation} by W.\ De Baere
of the paper:\\[3mm]
M.\ Renninger, ``Zum Wellen--Korpuskel--Dualismus'',
{\em Zeitschrift f\"ur Physik} {\bf 136}, 251-261 (1953).\\[1cm]
{\bf Translator's note:}\\[3mm]
In the original, german, version of the paper, the numbering of footnotes
and references
starts with 1 on each page. Because the text in the original paper and
in the translation, does not start and end with the same sentence
on each page,
we have chosen a consecutive numbering of the footnotes
and the references.\\[1.5mm]
It is the purpose of this translation, to bring Renninger's
important work
available to a broader audience in the field of the foundations of
quantum mechanics,
in particular to those concerned
with, and interested in, the significance and interpretation of the
quantummechanical wave function.\\[1cm]
{\bf Acknowledgment}\\[2mm]
We acknowledge the Editor of the former
{\em Zeitschrift f\"ur Physik}, Springer Verlag,
for giving us the permission to publish this english translation of
Renninger's original paper.

\newpage
{\large
\begin{center}
{\bf On Wave--Particle Duality}\footnote{This paper has been accepted
for publication, after the
correspondence of the author with
leading quantum physicists has convinced the Editors
of the fundamental importance of the subject.}
\end{center}
}

\vspace*{.5cm}
\begin{center}
M. Renninger\\
(Translated by W.\ De Baere,\\[1.5mm]
Laboratory for Theoretical Physics\\
Unit for Subatomic and Radiation Physics\\
Proeftuinstraat 86, B--9000 Ghent, Belgium\\
E--mail: willy.debaere@UGent.be)
\end{center}
\begin{center}
With 2 figures in text\\
(Received 5 june 1953.)\footnote{The manuscript was submitted
already on 10.4.53 to ``Naturwissenschaften'', but was
accepted for publication only later.}
\end{center}

\begin{quotation}
{\small
\noindent
{By means of a  thought--experiment,
consisting of an interference experiment with
two interfering beams,
it is shown that it can be demonstrated experimentally that
with {\em one single} particle
a wave can be associated which propagates in space and time as
a {\em physical reality}, i.e.\ that it
should not merely be considered as a distribution of probabilities.
The notion ``physical reality'' should be understood such
that, when this physical reality is considered in a
particular space at a particular time, it should be experimentally
possible to influence this reality in such a way that
future results of experiments show unambiguously that this
reality has been causally influenced by the experimental act
in this space and at that time.
}}
\end{quotation}

In modern physics it has become more and more the custom, to
discuss the last questions about quantum theory, especially the
wave--particle dualism, only by means of purely mathematical
considerations, and to consider the visualization,
or even the mere desire for it, as rudimentary and naive modesty.
It is, however, the purpose of the following discussion, to
point to some very precise conclusions, which follow merely
from purely experimental physical aspects, without any
previous knowledge of the mathematical quantum formalism,
which to my knowledge have never been obtained in this
way before.
At the same time the aim is to warn that in considering
such issues, the visualization
should not be given up {\em too early}, but should be kept on
as long as possible.
Clearly there is a limit in so doing, but
this limit should not be artificially set so as to exclude
interesting possible developments. It is precisely
the purpose of this work to present such a development.

In general one speaks of the two different ``pictures'' or
``aspects'', the ``wave''-- and ``particle--aspect''.
Detailed studies report the availability of a series
of experiments evidencing the wave--like nature, and another
series evidencing the particle--like nature of light and
matter.
\setcounter{footnote}{0}
\renewcommand{\thefootnote}{\arabic{footnote}}
But it has al\-ways been stressed, that there exists no
ex\-pe\-ri\-ment in
which {\em at the same time} both wave--like and particle--like
properties are observed, and in which the wave--like behaviour of
{\em one single} photon
or electron can be shown. According to P.\ Jordan\footnote{Jordan, P.:
Anschauliche Quantentheorie, p.\ 115. Berlin 1936.}
this were ``logically and mathematically nonsense''.
``It is impossible, that one single indivisible experimental act
shows up as well the one as the other appearance of light''.
Or as stated by
W.\ Heisenberg\footnote{Heisenberg, W.: Die
physikalischen Prinzipien der Quantentheorie, 2nd Ed., p.\ 7, 107.
Leipzig 1941.}
``Particle-- and wave--picture are two different appearances of one and
the same physical reality''. ``Now it is clear, that matter cannot
consist simultaneously of waves and particles \ldots'' ``The seeming
double nature has its origin in the inadequacy of our language''.
``Atomic processes don't have a visual representation.
Fortunately for the
mathematical description of these processes such a visualization is
not necessary altogether; a mathematical formalism of
quantum theory is at our disposal,
which accounts for all experiments \ldots''

In contrast with these statements I propose the following 3 propositions,
the proof of which will be given thereafter; they are first formulated for
light in the visual range, but they are also valid,
in appropriately modified form,
for radiation of other wavelength and for matter radiation:

1.\ It is possible to demonstrate experimentally,
that the {\em energy} of each photon
moves {\em across space and time} on a single continuous path,
and concentrated in
the form of a particle, i.e.\ at each instant it occupies a
connected region in
space\footnote{No specification is given
about the form and extension of this region of space, i.e.\ the
``particle''.}.

2.\ It can also be demonstrated experimentally, that
{\em with each single photon}
there may be associated a {\em guiding wave}
(without energy but ``causally'' influencable),
which obeys precisely the rules for the propagation of an
electromagnetic wave
(spatial extension, absorption, diffraction, interference,
splitting by reflection, refraction, double refraction etc.,
except energy),
hence has unambiguously a {\em wave nature}
which propagates likewise {\em in space and time},
i.e.\ as long as the photon is on its way it occupies at each instant a
specific region of space which is {\em not} necessarily single connected.

3.\ The connection {\em in space} of the particle of energy and of
the wave field is such, that the former can be found in a region
where the intensity
of the wave is different from zero, with a probability which is
proportional to this
intensity. Its propagation speed is equal to the group velocity
of the wave within
the limits set by Heisenberg's uncertainty relations.

Summarized briefly and drastically:
{\em Each quantum consists of a particle of energy,
``carried'' or ``guided'' by an energy free wave}.

This image is not new. It was already formulated by de Broglie in
1927\footnote{Broglie, L.\ de: Rapport au V.\ Congres Solvay
1927, p.\ 105.}
in terms of the by him introduced notion of ``pilot wave'',
but abandoned again
because of ``des objections tr\`es
graves''\footnote{Broglie, L.\ de: Ondes et Corpuscules, p.\ 34.
Paris 1930 and
Introdudions \`a l'Etude de la M\'ecanique Ondulation, p.\ 132.
Paris 1930.--
{\em Remark during the correction.} Only after submission of the
present manuscript,
the Editor became aware of the fact that de Broglie took up again
these former pictures since one year, in connection with relativistic
considerations, together
with J.-P.\ Vigier. In an article entitled: ``La Physique Quantique
restera--t--elle
Indeterministe?'' containing all his publications about this
issue so far, he
reports in an impressive way about his motivation, which lead him
to abandon 25 years ago his former concepts and to take up it
again presently. --
In the same way the Editor was not aware of the work by D.\ Bohm
[Phys.\ Rev.\ {\bf 85}, 166, 180(1952)],
in which de Broglie's concept of ``pilot--wave''
is developed theoretically, and which was the first main reason for de
Broglie to
develop further his former picture. \label{footnote2p253}}.
According to a citation from
N.\ Bohr\footnote{Schilpp, P.A.: Albert Einstein, Philosopher--Scientist,
p.\ 206. Evanston, Illinois 1949.\label{footnote3p253}},
also Einstein once speaks of ``ghost fields, guiding photons''.

On the other hand the 3 preceding
propositions are, however,
{\em neither intended as an explanation, nor as a mere visualization}.
Instead, both the energy particle and the guiding field associated with
the single photon are considered {\em simply as experimental facts}.
Attempts for explanation should follow {\em after} observation of the facts!

These statements will be proved by means of a
thought experiment, consisting of an elementary interference set--up.
In my opinion the
usual thought experiments for the analysis of the
wave--particle
duality (mostly refraction at a single or double slit) fail to penetrate
to the deepest possible level, and are not able to
reveal the most basic characteristics. Mostly they concern refraction
set--ups with a continuous intensity distribution. The main characteristics
of our set--up will be: {\em two} interfering rays, detection in
{\em two} possible places, {\em two} discrete intensities (probability
distributions) which may be chosen as either 1 or 0. This set--up is
shown in Fig.\ 1. (In an Appendix a more detailed description of the
set--up is given in order to prove its essential realizability.)

In the proposal it is assumed that two facts are experimentally proven:

1.\ The outcome of the interference experiment is the same whether many
photons contribute {\em in a short time} or in a long time: each
photon interferes {\em only with itself} (see e.g.\ Dirac,
Principles of Quantum Mechanics, p.\ 9).

\noindent
\begin{minipage}[b]{4cm}
\begin{center}
\epsfig{file=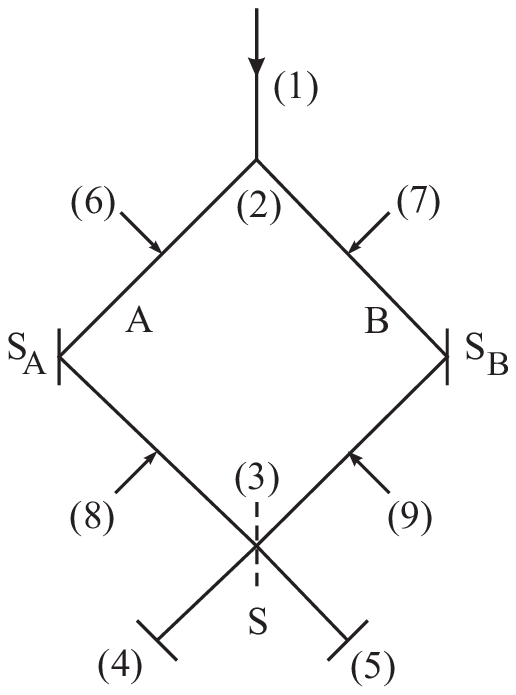,height=4.2cm,width=3.2cm}
\end{center}
\begin{minipage}[b]{4cm}
{\small Fig.\ 1.\ Interference set--up, schematic}
\end{minipage}
\end{minipage}
\hfill
\begin{minipage}[b]{10cm}
\vspace*{2mm}
2.\ Several coherent subrays of a light source {\em remain} coherent,
i.e.\ are able to interfere with each other after having traveled
arbitrary long distances along separated paths. This has been verified
experimentally up to distances of 2 km by Michelson and
Gale\footnotemark.\\
\hspace*{5mm}
A parallel ray (1) is split by (2) in 2 rays $A$ and $B$ which move, in
separate spaces, through a system of pipes towards (3). There they
interfere with each other, e.g.\ after transmission through and reflexion
by a beamsplitter $S$, and arrive in places (4) and (5) where they can be
detected. \label{page254}
\end{minipage}
\footnotetext{Michelson and Gale: Nature, Lond.\ {\bf 115}, 556 (1925).}\\[4mm]
The path difference between $A$ and $B$ can be set up (e.g.\ by displacing
one of the mirrors $S_A$ or $S_B$) such that it is either + or
$-\lambda/4$, resulting either in brightness in (4) and darkness in (5),
or the other way round (the ray which is reflected by the beam splitter
$S$ gets a phase shift of $\pi/2$, the phase of the transmitted ray
remains unchanged).

The light source should:

1.\ Allow the transition to
such weak intensities, that never (i.e.\ with a probability
tending to zero), more than one photon at the same time is present
between (2) and (3).

2.\ Each time inform the observer by means of a signal the instant
-- taken each time as $t=0$ --, when a photon has passed the
point (2), within an uncertainty $\pm \Delta t$,
which is small compared with the time of passage between
(2) and (3)\footnote{With respect to this point, see the Appendix
for the essential possibility of such a signal.}.

The perfection of the optical set--up [parallelism of the ray, reflection
and precision of the mirrors, perfection of the detectors in (4) and (5)
etc.] should be such that each photon that passes (2) will be observed
either in (4) or in (5) (e.g.\ by means of scintillation or phototube)
with a probability very close to 1 (i.e.\ almost
with certainty). Thus if the path
difference between $A$ and $B$ is set up such that brightness will
be observed in (4), darkness in (5), then each photon entering (1) will be
observed in (4).

At different places, such as (6), (7), (8) and (9) it should be possible,
to insert in the paths $A$ and $B$ a slide, either an opaque screen with
detector, which -- as in (4) and (5) -- registrates each photon which is
absorbed by the screen, or a $\lambda/2$ plate. The paths $A$ and $B$
should be long enough (say some light--minutes), in order to allow the
insertion of such slides on different instants while a photon is moving
between (2) and
(3)\footnote{This does not require large signal speeds, because the
paths $A$ and $B$ can be deflected as many times as necessary, so that
the positions (2) to (9) are at the same time separated in space
and in the immediate neighbourhood of the observer.}.

With this set--up we make then the following observations
[Reduced intensity as mentioned above,
path lengths adjusted so that
brightness is observed in (4), darkness in (5)]:

a) As long as nothing is inserted in the places (6), (7), (8) and (9),
each photon passing (1) will be observed in (4), nothing in (5).

b) If in (6) a slide is inserted within the interval
between $t_6-\Delta t$ and $t_6+\Delta t$,
where $t_6$ is the time the photon needs to
move from (2) to (6), $t_6=l_6/c$ [$l_6$= distance between (2) and
(6)], then for the observation there are two possibilities,
each of which is realized with probability 50\%:

$\alpha$) The photon is detected in (6). It is then absorbed
and has disappeared, and no further experimental intervention
can result in a further observation, e.g.\ revealing the
existence in path $B$ of a further existing physical reality.

$\beta$) The photon is {\em not} detected in (6). Then
it can be experimentally verified that with certainty it can
be detected e.g.\ either in (9) at
the time $t_9=l_9/c$, or, when this is not the case, to
detect it {\em either} in (4) {\em or} in (5), with
probability $\frac{1}{2}$.
Blocking path $A$ has, when the quantum is not observed in it,
yet a definite physical effect, namely that there, where in the
case of both open paths no quantum may be observed, now a quarter
of all entering quanta will be detected; and it is indifferent,
whether path $A$ or path $B$ is closed.

c) Insert now in (6) a fully transparent $\lambda/2$ plate instead
of an obsorbing slide, again within the time interval
$t\pm \Delta t$. The result is:
{\em All photons are now detected in (4) instead of in (5)}.
The same result is obtained when the plate is inserted in (7) instead
of in (6). This means, that it is possible, by this experimental
intervention, to send the photon from (4) to (5),
{\em at will from within each of the paths $A$ and $B$},
i.e.\ to influence it {\em causally}. Yes, it is possible,
to ``cancel'' an already decided ``bending'', by inserting
at corresponding later times  $t_8$ resp.\ $t_9$ a further --
or also the same -- $\lambda/2$ plate.

With these results the 3 abovementioned
propositions are proven:

Proposition 1 by experiment b. The photon can always be
absorbed, i.e.\ detected, only in {\em one} of the paths
$A$ and $B$, never in both together, and further each time
at a specific instant, determined by the propagation speed.
The fact, that the {\em certainty}, to ``find'' the photon in one
of both paths -- and when it is not found in {\em one} path, the
certainty to find it {\em then} on a corresponding later
instant in the {\em other}
one -- moves with speed $c$, leads to the compelling
conclusion, that {\em the photon itself} (i.e.\ its total energy)
moves inside a confined self contained region on a
{\em continuous path} through space and time.
The conclusion of the localization of a {\em system} in a region,
from the {\em certainty} to {\em detect} that system in that region, is
undoubtedly allowed, and even required by the most elementary logic.

The validity of Proposition 2 follows from observation b, $\beta$
and especially from c. By the intervention of inserting the phase
plate in one of both paths, whatever one, the single photon can be
steered at will between both possible detection places. This proves,
that with each single photon there is associated a
{\em physical reality which simultaneously moves on both paths}.
Additional observations, which will not be discussed further, show,
that the evolution of this ``physical reality'' obeys in all points
the laws for the propagation of an {\em electromagnetic wave},
except the fact, proven by observation b, that it does not contain
energy (the total energy moves on {\em one} of both paths;
because the wave moves on {\em both} paths, it does not contain energy).

Proposition 3, which basically should be considered as a supplementary
one, is proven by the combination of all observations, but in
particular by the fact, that the experimental interventions of inserting
slides or phase plates has the described result, when they happen
within the mentioned time interval,
which must be the same in each path, independent whether
it is an intervention on the energy particle (by means of slides) or an
intervention on the wave (by means of phase plates).

Are now, after all these considerations, the wave--``picture'' and
particle--``picture'' really but ``two forms of appearance of one and
the same physical
reality''? Is it not, on the contrary, the case
that wave and particle
are {\em two separate physical realities} associated with single photons?
Did the observation c above not ``reveal in one single indivisible
experimental intervention both the one and the other picture of the
nature of light''? (The result of inserting the phase plate proves
the wave nature, the observation in one single point proves the
particle nature of one and the same photon).

Of course one is free, to speak of the wave as a pure
``probability''--wave. But one should be aware of the fact,
that this probability wave propagates in space and time in a continuous
way, and in a way that she can be influenced in a finite region of space
-- and only there! -- and also at that time! --, with an unambiguous
observable physical effect!

I am not sure, whether the facts and conclusions discussed in this work
have some
connection with the doubts of A.\ Einstein about the completeness
of the description of elementary processes by quantum
mechanics\footnote{Summarized in: P.A.\ Schilpp, Albert Einstein, see
footnote 6 page \pageref{footnote3p253}.}.
It is obvious, that precisely the proven reality of the wave associated
with the single particle, which quantum mechanics,
as it is stated explicitly, is unable to account for,
may be considered precisely as an expression of the
incompleteness maintained by
Einstein\footnote{{\em Remark during the correction}.
As the articles which the Editor became aware of since the submission
of the manuscript (see footnote 5 p.\ \pageref{footnote2p253})
show, and in accordance with
a kind personal anouncement by Mr.\ Einstein this is indeed the case.
In this respect also the work by L.\ J\'anossy
[Ann.\ Phys.\ {\bf 11}, 323 (1953)] should be mentioned, which attempts
to give an alternative deterministic account of quantummechanical
processes, although there is a violation of the assumption of
the coherence of separated rays over arbitrary
distances, the validity of which we accepted in our work
(see page \pageref{footnote3p253}).}.
In any case was our definition of ``physical reality'' inspired by
the definition of Einstein, Podolsky and
Rosen\footnote{Einstein, A., B.\ Podolsky and N.\ Rosen:
Phys.\ Rev.\ {\bf 47}, 777 (1935).}
of the same notion.

In the same way I cannot say, whether there is a connection with the
new work by
W.\ Weizel\footnote{Weizel, W.: Z.\ Physik {\bf 134}, 264 (1953).},
in which it is tried to refute von Neumann's proof of the
impossibility of a causal model for quantumtheoretical processes.

For sure, the picture of a particle guided by a wave devoid of energy,
which is intended to be more than a ``picture'', namely ``reality'',
nowhere contradicts quantum mechanics, and may moreover be a
valuable aid for the visual comprehension of elementary processes and
for making exact prognoses about the outcome of experiments.

Heisenberg's uncertainty relation,
as far as it has something to do
with the motion of a particle as treated here,
may be given also a visual significance:
it states, that after emission of the particle it is impossible
to know where, i.e.\ {\em at which point in the wave} the particle,
the knot is situated [as well transversally, i.e.\
{\em in which direction}, as longitudinally,
i.e.\ {\em the distance} within the coherence length $\Delta x$
determined by the uncertainty relation
(the expression of the uncertainty relation for light is:
$\Delta x\cdot \Delta 1/\lambda\geq 1$).
Yet, this relation says nothing about
our additional statement, which restricts the uncertainty,
that within the space made available by the uncertainty relation,
the particles follow a continuous path in space and time,
and that once a certain direction is taken, it keeps moving in
this direction. The ``coordinates'' of the particle within
the guiding wave should be considered as ``hidden parameters'' in
v.\ Neumann's sense.

It is clear that the concept of a ``wave devoid of energy''
is disturbing. However, in attempting to avoid this one should
realize that the experimental results
would allow only much less attractive alternatives:
because the existence of the wave can be experimentally
verified during the entire motion of the photon,
and the existence of the energy only in two points,
the emission and the absorption point, the assumption
of the existence of a normal electromagnetic wave
containing energy would have the
unavoidable consequence, that at the moment of absorption
the wave would contract with superluminal speed, and moreover
through closed walls. Such assumption would be completely
unacceptable.

The author is fully aware of the fact, that the uneasiness of the
idea of a wave devoid of energy still increases, when the
consequences are considered, of the establishment that
the laws of propagation of this wave are in every respect
those of an electromagnetic wave. For instance there is
refraction
(which in the Maxwell--Lorentz theory has
its origin in
energy exchanges with dipole oscillations of the electrons),
double refraction, polarization, scattering etc.; this are
all manifestations which come about through interactions with matter.
Could this uneasiness be relaxed by the consideration, that
also matter must have guiding waves, and that these are
the ``spooky'' entities, to use Einstein's expression,
similar with ``matter waves''?
That the guiding waves ``{\em devoid of energy}'' associated
with {\em energy radiation} should correspond with
guiding waves {\em devoid of matter} associated with
{\em matter radiation}? And that the interactions between
these spooky entities should co--determine the events
resultant from the {\em energy} interactions between
the lumps of energy, resp.\ matter?
A directly unobservable world behind the observable one,
guided by the former? But, to stress it again, a world the
reality of which can
in time and (three dimensional) space be experimentally
followed up, and which does not exist only in the form of abstract
probabilities\footnote{See footnote 15, page  \pageref{footnote1p261}.}.

What happens to the wave devoid of energy of a photon after its
absorption? When it is absorbed for example in (6), and when
in addition the detectors in (4) and (5) are removed, what
happens then with the wave in $B$? Does she move further
towards infinity, or does she disappear at the moment of absorption?
Of course this question cannot be answered principally.
The former assumption appears to be the more natural one, because it
avoids the conclusion to the existence of influences which propagate
with infinite speed also through closed walls, a conclusion which
within the physical world is inconceivable. In any case were such
influences not associated with transport of energy.

It is still possible to consider the following issue:
what happens to the particle when the wave is splitted by
reflection at a refracting surface, or at a partially transparent
mirror, or by double refraction, i.e.\ each time a wave is splitted
in more component waves with different propagation vector and mostly
also different polarization state? All these cases can be understood
without contradiction with the help of the plausible assumption,
that each time the particle follows one of the component waves,
with a probability which is proportional to the
intensity of this component wave, in the same way as a particle
in a splitting fluid current. It is not necessary to
associate with the particle properties, such as polarization.
These should be associated only with the guiding wave.

\begin{center}
{\bf Appendix.}
\end{center}
In order to anticipate objections against the {\em principal}
(not the {\em practical}) feasibility of the described
experiments, more precise technical details will be given
which support their realizability.
The set--up should have the
characteristics as shown in Fig.\ 2:

An almost pointlike monochromatic light source $L$ is situated at
the center of a small half spherical convex mirror $S_1$,
which itself is large as compared to $L$. Furthermore $L$
is in the focus of a parabolic mirror $S_2$ which again is
large with respect to $S_1$. In this way the spherical wave
originating in $L$ is transformed in a parallel bundle (1),
which contains the photons.
\begin{minipage}[b]{\linewidth}
\begin{center}
\epsfig{file=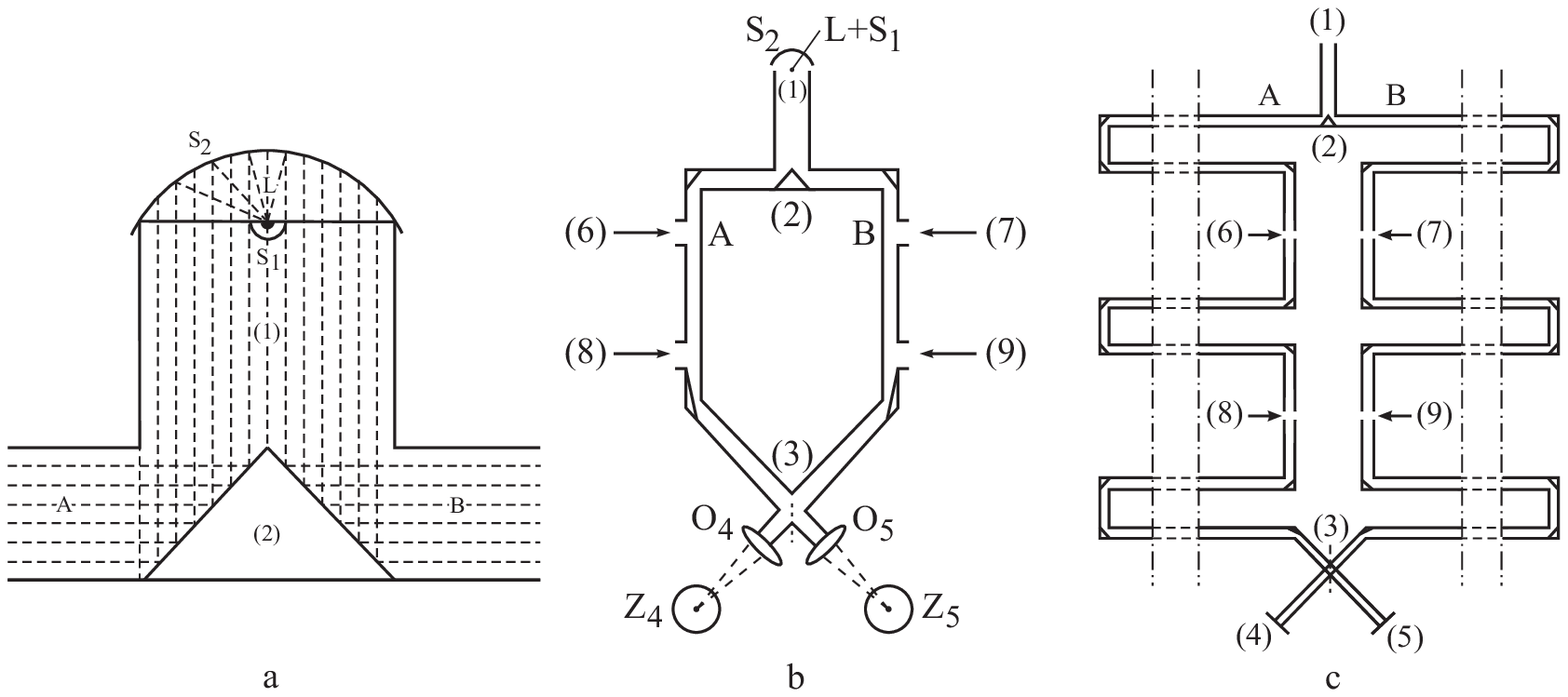,height=6cm,width=\linewidth}
\end{center}
\begin{minipage}[b]{\linewidth}
{\small Fig.\ 2a--c. Technical details of the set--up.
a Light source with reflecting mirrors; b Complete set--up;
c Bringing together paths A and B through multiple bending.}
\end{minipage}
\end{minipage}

\vspace*{3mm}
This ray (1) is split at (2)
in 2 separated rays $A$ and $B$ by means of 2 mirrors
which are inclined by 45\textdegree. In order to remove
the negative effects of the angular divergence of the parallel rays
(due to the finite dimension of $L$) the parabolic mirror may have
a large section, i.e.\ the focus and diameter can be
chosen arbitrary large.
The following choices may be made:
diameter of the light source: $10^{-6}$ (100 \r{A}),
$S_1$: 1 cm, $S_2$: $10^6$ cm (10 km), total light distance
between (2) and (3): $10^{12}$ cm (30 light--seconds).
As already mentioned, for performing experiments b and c it is
required that the places (2) to (9) are close to each other.
Therefore, the separated paths must come
repeatedly in the neighbourhood of the observer, which is
illustrated in Fig.\ 2c. In (6), (7), (8) and (9) the section of the
rays may be decreased by optical arrangements, in order to
facilitate the insertion of slides and phase plates. -- Both
lenses $O_4$ and $O_5$ focus the outgoing rays (4) and (5)
on the ingoing gates of two photomultiplicators. The detectors to be
inserted in (6), (7), (8) and (9) equally consist of such aggregates
(lenses and counters).

An indication is still needed how the observer may get information
of the time of emission of a photon within an uncertainty interval
which is small compared to the time needed by the photon for
moving between (2) and (3).
This could be resolved as follows: The light source $L$ consists of
a small hollow sphere, which contains a small amount of gas the
atoms of which have for simplicity only one excitation level above the
ground level. A beam of electrons with known velocity
is sent through this sphere containing the gas,
and with such an intensity
that only one atom is excited in a time interval which is large
compared with the time needed for a photon to move between (2) and (3),
in our example of the order of hours. A supplementary arrangement
measures the energy of the transmitted electrons and informs the
observer as soon as an electron is detected which has lost
an energy corresponding to the excitation energy of the atoms.
This signal warns the observer, that one of the atoms is excited
and a photon will be emitted in the apparatus within the time needed
to return to the ground state (of the order of $10^{-8}$ sec).

The first compositions of the above ideas date already from
20 years ago. The final explanation of the present
formulation benefited in a substantial way from a series
of very exciting discussions with Dr.\ S.N.\ Bagghi and
R.\ Hosemann\footnote{{\em Remark during the correction}.
In a work in preparation these authors investigate the present
issue in terms of their newly introduced ``Algebra physikalisch
beobachtbarer Funktionen mittels Faltungsoperationen'',
Part I. [Z.\ Physik {\bf 135}, 50 (1953)],
and come, following a kind personal announcement, to
a confirmation of the guiding wave picture, in which
the amplitude of the guiding wave has the dimension of energy.
In fact, the average {\em in time and in space} of the energy density
of such a wave should be zero at any time and in each point.
How far the
related assumption of ``negative energy densities'' of, resp.\
the necessity to atribute a kind of ``level of energy density''
to, empty space can be carried through, will not be discussed here.
\label{footnote1p261}},
Fritz Haber--Institute, Berlin--Dahlem, in the course of the
last half year.

{\em Marburg a.d.\ Lahn}, Crystallographic Institute of the
University.\\[1cm]

\end{document}